\documentclass[11pt]{article}
\usepackage{moriond,epsfig}

\usepackage{subfigure}

\def\Journal#1#2#3#4{{#1} {\bf #2}, #3 (#4)}

\def\PRL{\em Phys. Rev. Lett.}
\def\PRD{{\em Phys. Rev.} D}

\def\be{\begin{equation}}
\def\ee{\end{equation}}
\def\bea{\begin{eqnarray}}
\def\eea{\end{eqnarray}}

\begin{document}
\vspace*{4cm}
\title{Di-Photon and Photon+b/c Production Cross Sections at Ecm=1.96 TeV}

\author{Anant Gajjar (on behalf of the CDF Collaboration)}

\address{Department of Physics, University of Liverpool, \\
Liverpool L69 7ZE, England}

\maketitle\abstracts{
Measurements of the di-photon cross section have been made in the central region and are found to be in good agreement with NLO QCD predictions. The cross section of events containing a photon and additional heavy flavour jet have also been measured, as well as the ratio of photon+b to photon+c. The statistically limited sample shows good agreement with Leading Order predictions. }

\section{Introduction}
\label{sec:intro}
We present results for the inclusive di-photon cross section\cite{dipho} and the photon plus heavy flavour analyses using data taken by the CDF collaboration. The measurements provide an understanding of QCD production mechanisms and are also possible signatures for ``new'' physics. The di-photon analysis is sensitive to initial state soft gluon radiation and can be used to measure gluon PDFs. It is also one of the main backgrounds for Higgs to $\gamma\gamma$ searches that will be performed at the LHC. The ratio of  photon + b to photon + c is sensitive to the charm content of the proton.

\section{Production Mechanisms}
\label{sec:prod}
\begin{figure}
\begin{center}
\subfigure[Direct]{\epsfig{figure=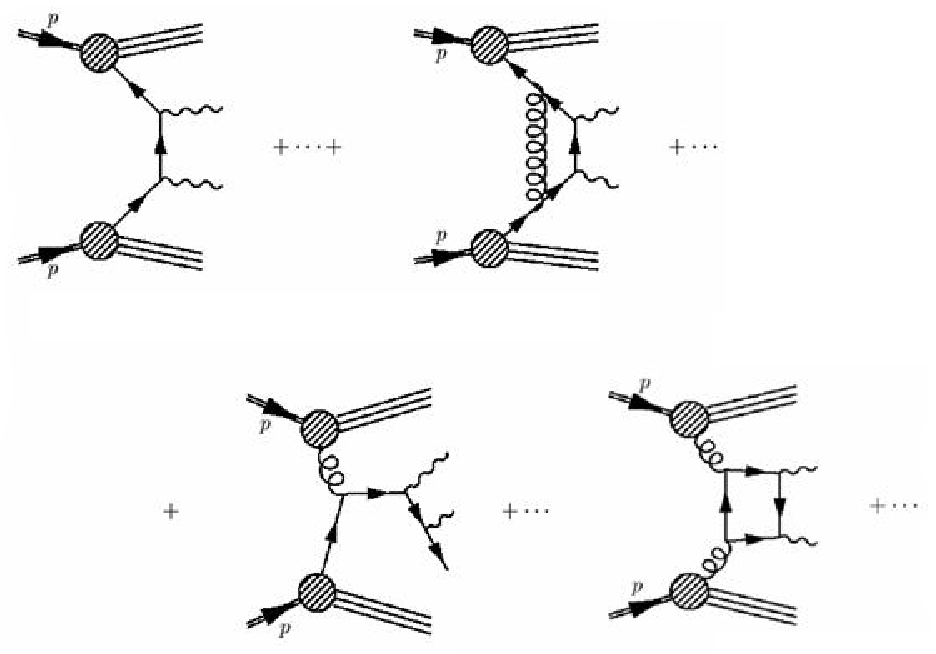,height=1.5in}\label{fig:dir}}
\subfigure[Fragmentation]{\epsfig{figure=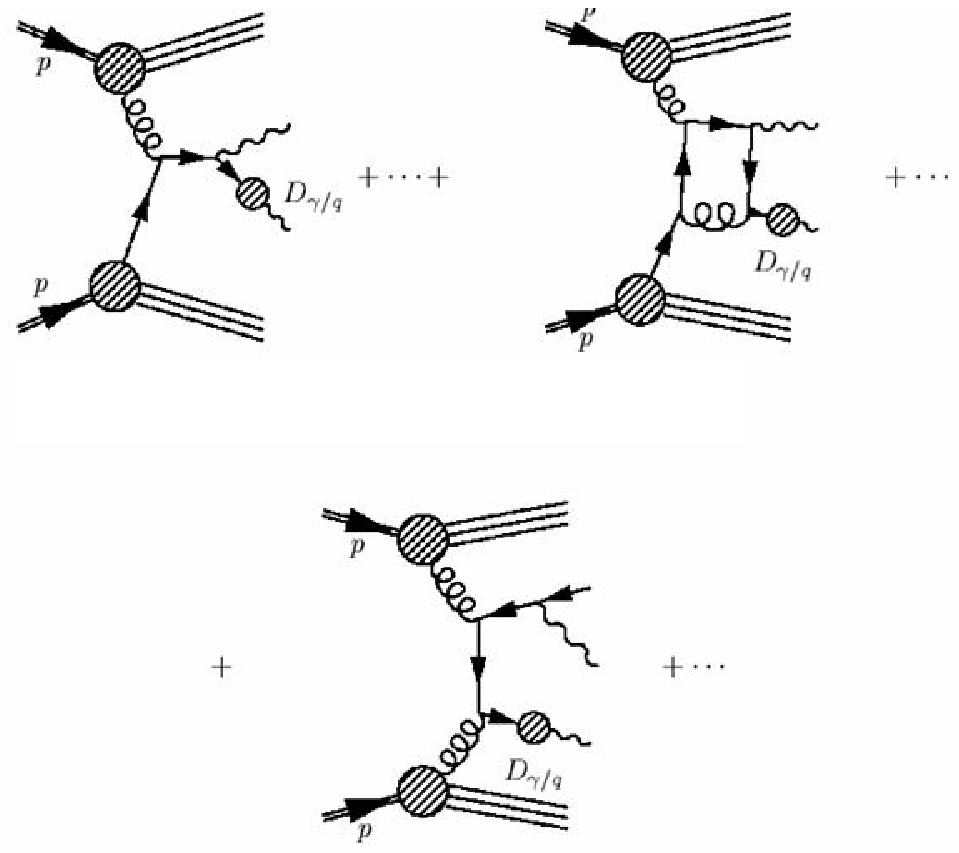,height=1.5in}\label{fig:frag}}
\end{center}
\caption{Feynman diagrams showing prompt di-photon production mechanisms through a) direct processes and b)parton fragmentation processes.
\label{fig:prod}}
\end{figure}

Prompt di-photon production can be split into two categories: direct processes where the photons are produced directly through $q\bar{q}$, $qg$ and $gg$ interactions, shown in figure \ref{fig:dir}; and fragmentation processes where at least one of the two photons is produced through the fragmentation of a parton, shown in figure \ref{fig:frag}. Both processes contribute to Leading Order and Next to Leading Order calculations.  The experimental results for the di-photon analysis are compared to three predictions: DIPHOX\cite{diphox}, ResBos\cite{resbos} and PYTHIA\cite{pythia}. DIPHOX contains all processes at Next to Leading Order\footnote{The gluon-gluon process is only available at Leading Order in DIPHOX, however recent calculations \cite{ggnlo} have been added to the DIPHOX results.}. ResBos contains all direct process at Next to Leading Order, Leading Order fragmentation processes and includes soft gluon resummations. Pythia includes all diagrams at Leading Order.

Photon plus heavy flavour events are produced by direct radiation of a photon from the heavy quark, and also when a photon is radiated off a light quark and the heavy flavour quarks are produced through gluon splitting. Leading Order Pythia predictions for these processes are used to compare the experimental results to.

\section{Photon Identification at CDF}
\label{sec:photon}
The Central Electromagnetic Calorimeter (CEM) and two wire chambers, the Central PreRadiator detector (CPR), located just in front of the CEM, and the Central Shower Maximum detector (CES), embedded inside the CEM at 6 radiation lengths are used to identify photons in the central region. Photon candidates are required to have an isolated\footnote{The transverse energy within a cone of radius 0.4 around the photon is required to be less than 1GeV} electromagnetic shower with no associated track. The energy deposited in the hadronic calorimeter is required to be very small in comparison to the electromagnetic energy.

Neutral mesons decaying to multiple photons are the main background to prompt photon identification. Two statistical techniques, the CES method and CPR method, are used to estimate this background. The CES method uses the shower profile, which is different for multiple photons compared to single photons, to estimate the number of ``fake'' photons. The method is only valid for photons with a transverse energy, $E_t$, less than 35GeV. Above this the multiple photons from neutral meson decay are collinear and thus indistinguishable from single photons. The CPR method uses the fact that the probability of a conversion is higher for multiple photons than for a single photon. As a result multiple photons are more likely to generate a hit in the CPR, which identifies conversion electrons through hits in the detector. 

\section{Di-Photon Production}
\label{sec:di-photon}
A photon triggered dataset with an integrated luminosity of $207 pb^{-1}$ is used to measure the cross section. Candidate events are required to have two photon candidates within a pseudo-rapidity range, $|\eta| <0.9$, where one of the photons has $E_t > 13GeV$ and the other has $E_t >14GeV$. The number of true di-photon events is derived using the CPR and CES methods to estimate fake photon content.

\begin{figure} [htb!]
\begin{center}
\subfigure[$\frac{d\sigma}{dM_{\gamma\gamma}}$]{\epsfig{figure=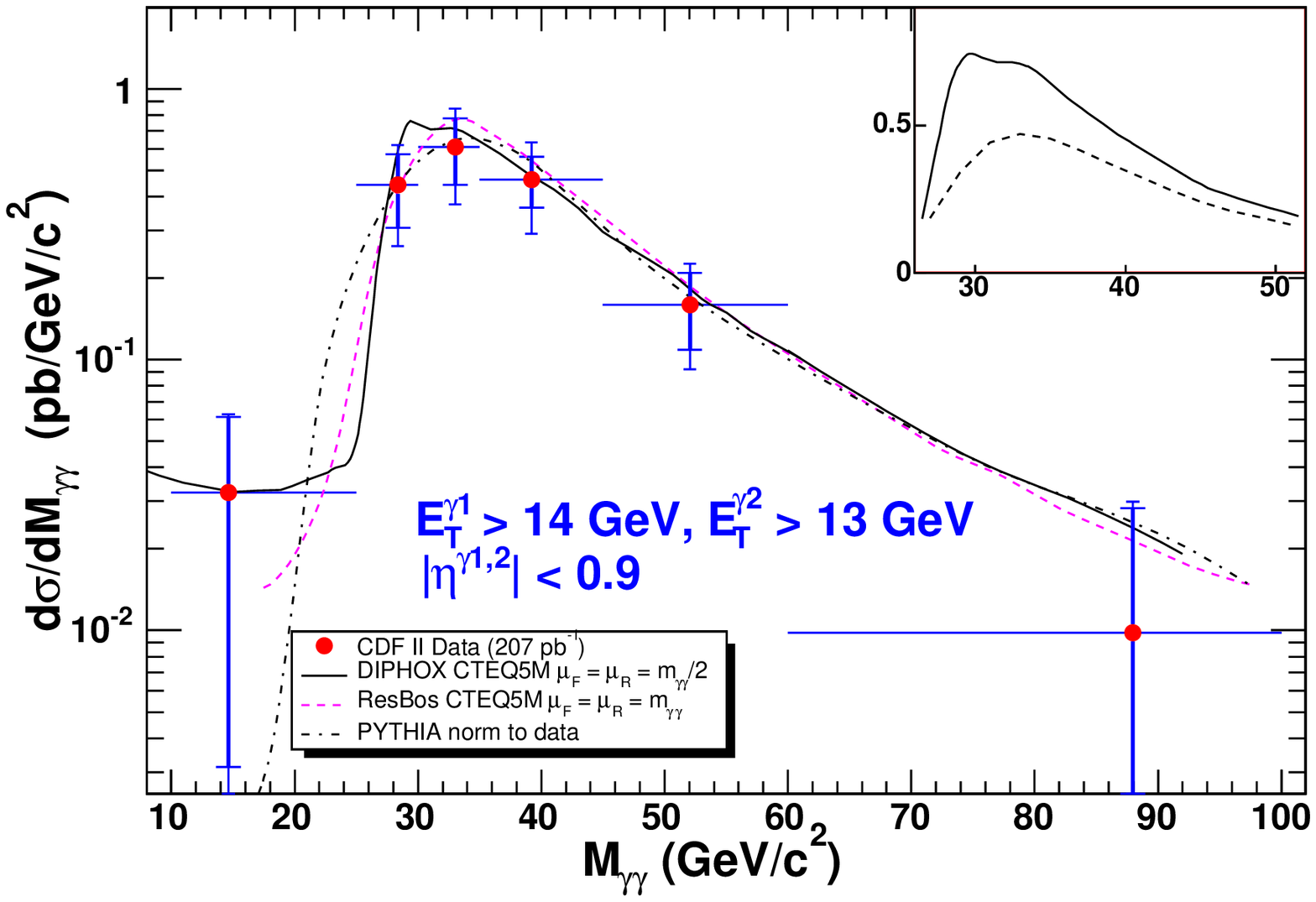,height=1.5in}\label{fig:mass}}
\subfigure[$\frac{d\sigma}{d\Delta\phi_{\gamma\gamma}}$]{\epsfig{figure=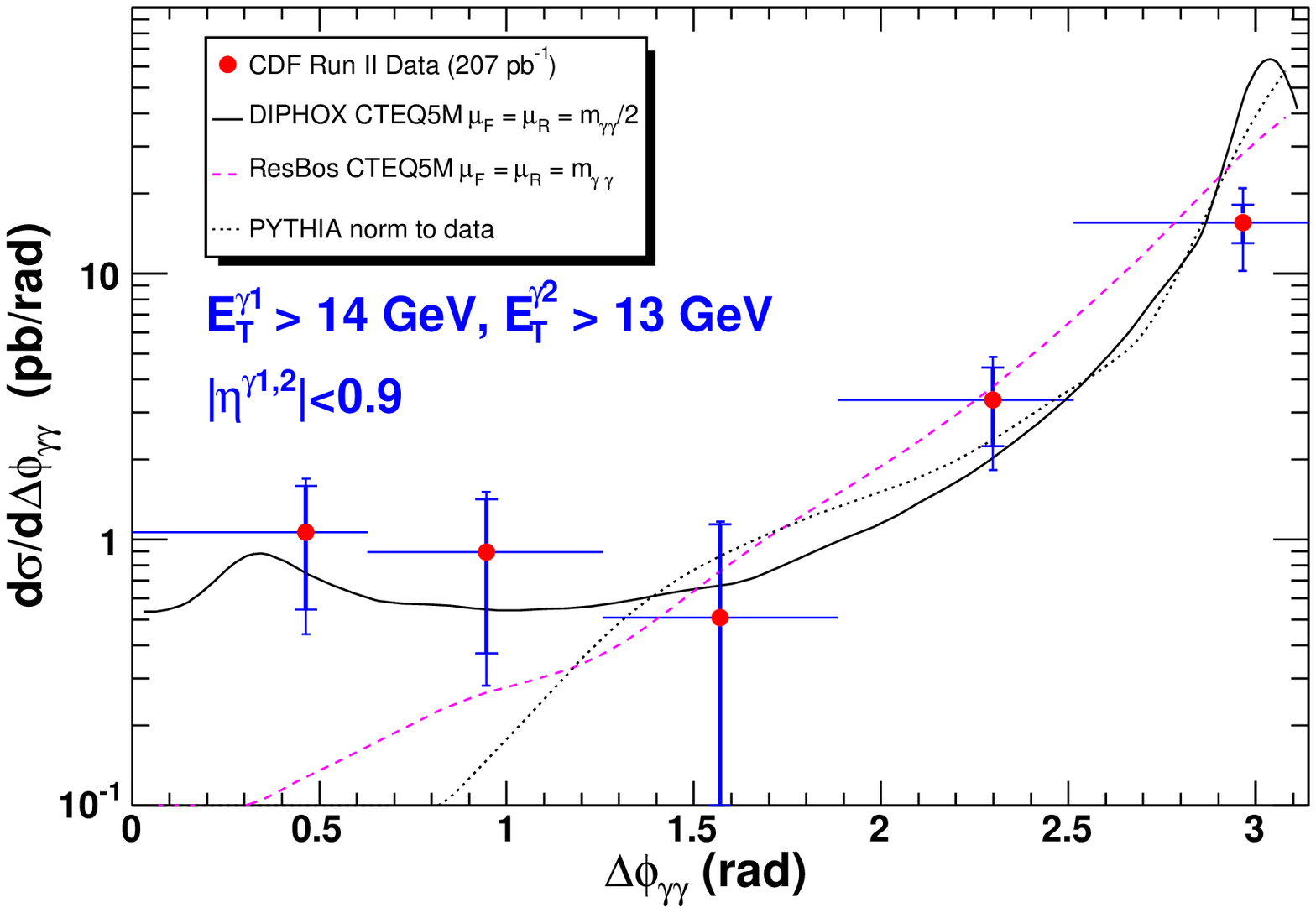,height=1.5in}\label{fig:dphi}}
\subfigure[$\frac{d\sigma}{dq_T}$]{\epsfig{figure=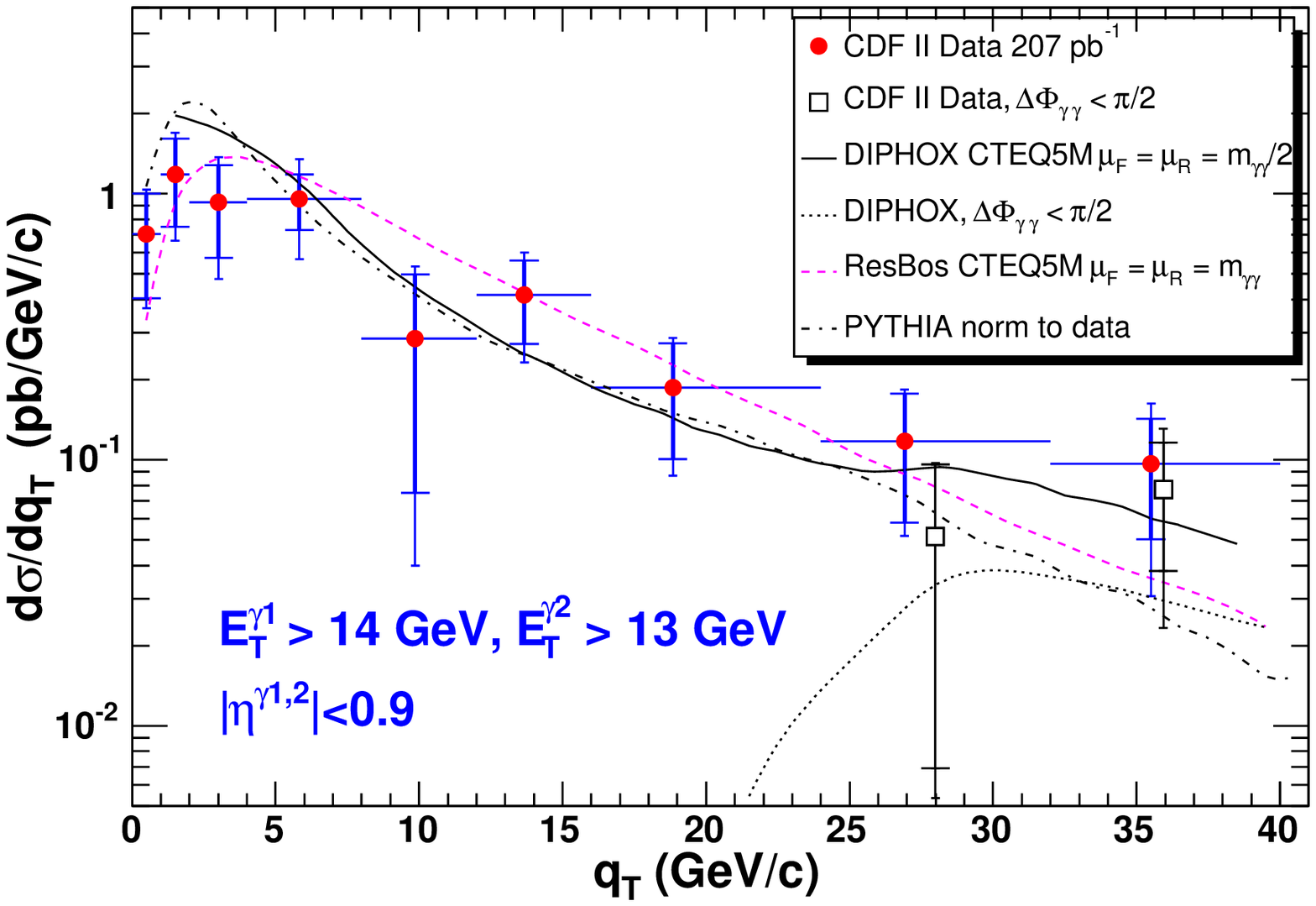,height=1.5in}\label{fig:qt}}
\end{center}
\caption{Di-Photon differential cross section as a function of: a) the invariant mass of the two photons, where the inset shows the DIPHOX prediction with (solid) and without(dashed) the gluon-gluon predictions; b) the azimuthal angle between the two photons; c) the transverse momentum of the two photon system. The Pythia predictions are scaled by a factor of 2 in all plots.}
\end{figure}

Figure \ref{fig:mass} shows the di-photon differential cross section as a function of the invariant mass of the two photons. Data, DIPHOX and ResBos agree well at high invariant mass. Data appears to be closer to the DIPHOX prediction (which includes all diagrams at NLO) at low invariant mass, however the statistical error is large. The Pythia predictions are too low  and need to be scaled up by a factor of 2 for agreement, although the shape of the spectrum is in good agreement. The inset shows the DIPHOX prediction with and without the gluon-gluon contribution. It highlights the importance of the gluon-gluon contribution at low invariant mass, where the prediction is considerably higher.   

The differential cross section as a function of the azimuthal angle between the two photons is shown in figure \ref{fig:dphi}. The region $\Delta\phi < \pi/2$ is only accessible at NLO. In this region data shows good agreement with DIPHOX predictions. There is better agreement with ResBos at larger opening angles as the gluon resummations contribute more in these regions. Again the Pythia predictions are not in agreement.  

The differential cross section with respect to the transverse mass of the two photon system ($q_T$) is shown in figure \ref{fig:qt}. In the low $q_T$ region the DIPHOX calculation is unstable, as the NLO order calculation is divergent. The soft gluon resummations in ResBos are required for a physical result and show good agreement with data in this region. At high $q_T$ the region of phase $\Delta\phi < \pi/2$ is accessible at Next to Leading Order, which results in a shoulder in the DIPHOX calculation. This is not seen in ResBos as the fragmentation processes are only calculated at Leading Order. The data suggests better agreement with DIPHOX, but is statistically limited. 

\section{Photon + Heavy Flavour Production}
\label{sec:phot_heavy}
A photon triggered dataset corresponding to an integrated luminosity of $67pb^{-1}$ is used for this measurement. A photon candidate with $E_t>25$ and a jet containing a secondary vertex is required. The composition of the sample is found by fitting secondary vertex mass templates for b, c and light quark jets, derived from Monte Carlo, to the secondary vertex mass spectrum in data. The photon background is estimated using the CPR method and the cross section found as a function of the photon $E_t$.

\begin{figure}[htb!]
\begin{center}
\subfigure[$\gamma+b$]{\epsfig{figure=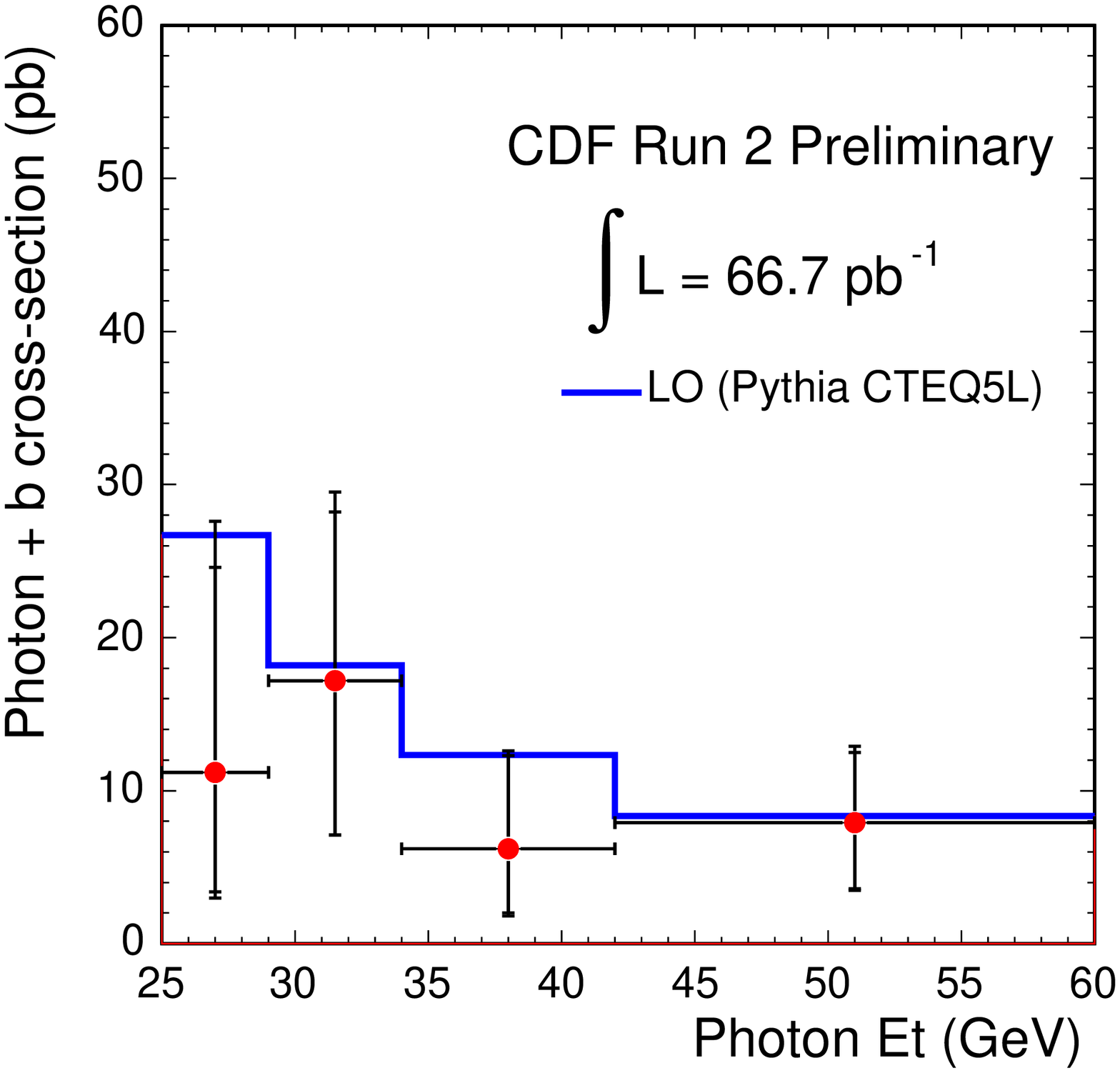,height=1.5in,angle=-90}\label{fig:gamb}}
\subfigure[$\gamma+c$]{\epsfig{figure=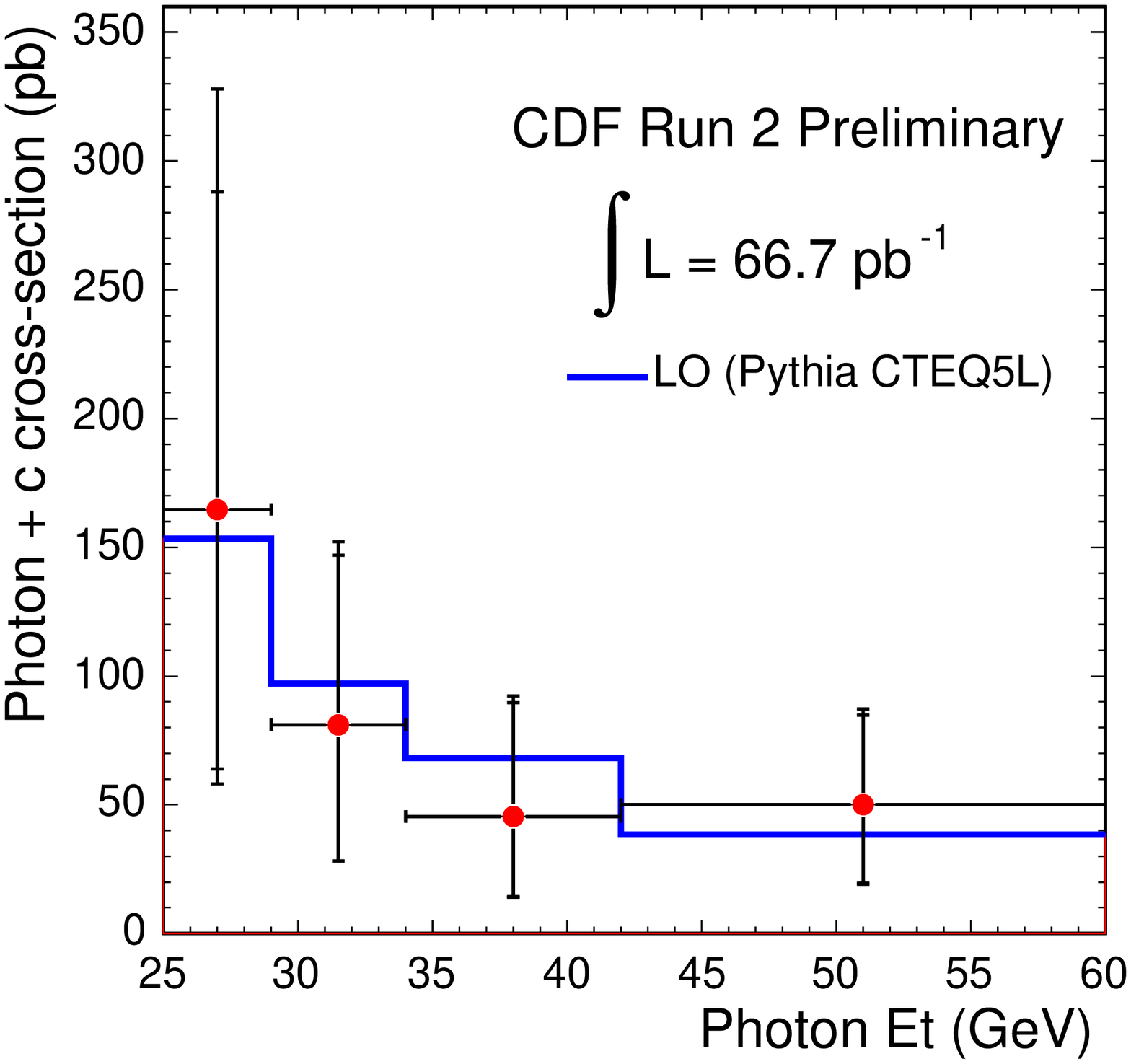,height=1.5in,angle=-90}\label{fig:gamc}}
\subfigure[$\frac{\gamma+c}{\gamma+b}$]{\epsfig{figure=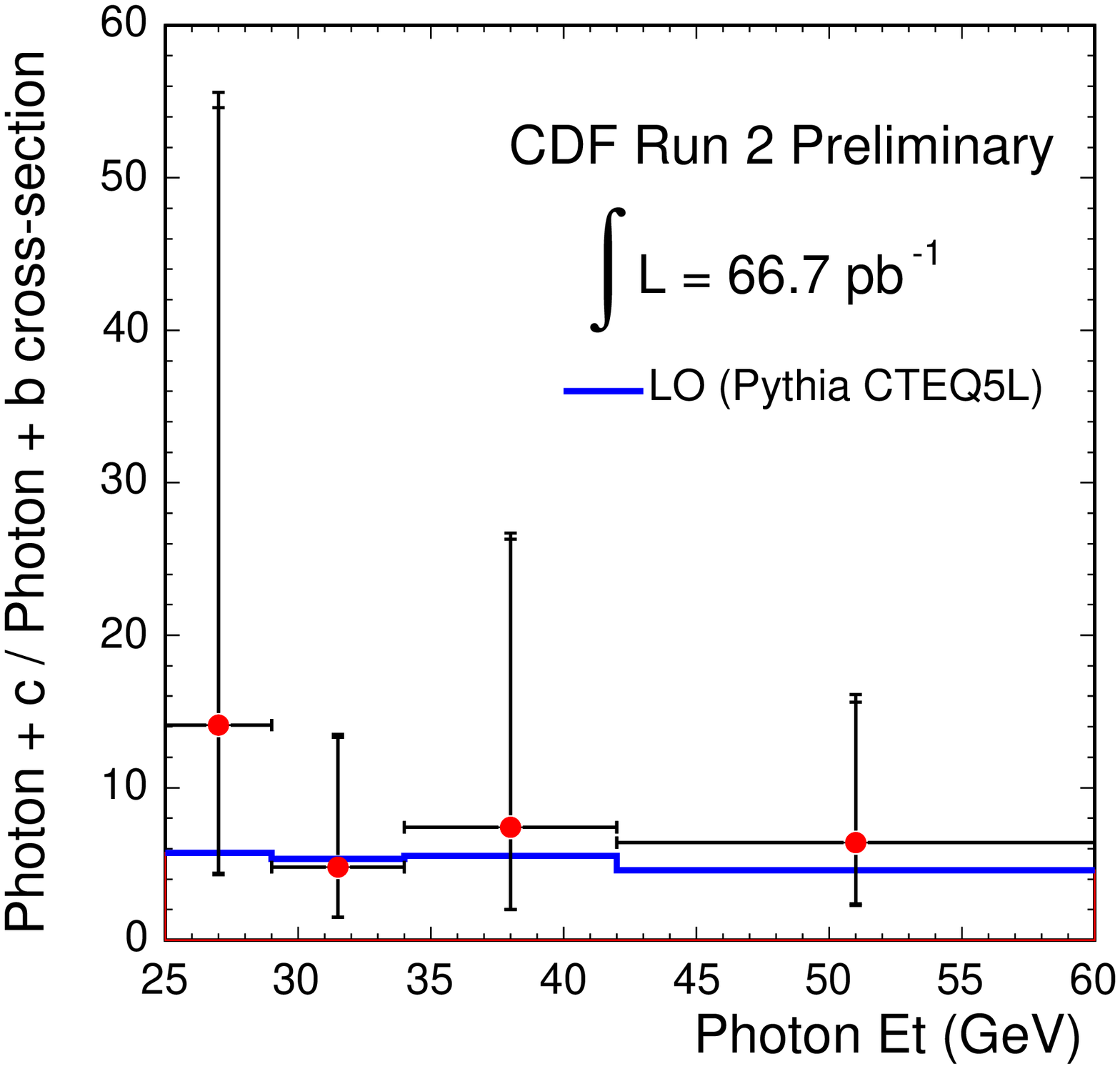,height=1.5in, angle=-90}\label{fig:ratio}}
\end{center}
\caption{Cross Section as a function of photon $E_t$ for: a)photon + b production; b) photon + c production; c) ratio of the photon + c cross section to the photon + b cross section.}
\end{figure}

Figures \ref{fig:gamb} and \ref{fig:gamc}  show the photon + b and photon + c cross sections, respectively, as a function of photon $E_t$. The Leading Order Pythia predictions are in good agreement with data. The ratio of photon + c to photon +b, shown is figure \ref{fig:ratio}, also shows good agreement between data and Pythia predictions but have a large statistical uncertainty. 

\section{Conclusions}
\label{sec:conclusions}
The di-photon production rate has been measured at CDF and shows good agreement with Next to Leading Order and resummed calculations in different regions of phase space. A full Next to Leading Order Calculation including soft gluon resummations is necessary to see agreement in all regions. CDF is working on improving the analysis by extending to the forward regions and including the additional data that has been collected. The results show there is sensitivity to gluon-gluon interactions, and we aim to measure gluon PDFs.

The photon + heavy flavour cross sections have also been measured at CDF and show agreement with Leading Order Pythia predictions. The analysis is statistically limited. Work is ongoing to include more data and compare results to Next to Leading Order Predictions, with the aim of presenting new results, together with limits on exotic processes, in the near future.

\end{document}